\documentclass[preprint, amsmath,amssymb]{revtex4}
\usepackage{graphicx}
\pdfoutput=1
\begin{document}
\title{Simulation and Analysis of Superconducting Traveling-Wave Parametric Amplifiers}
\thanks{This article will appear in \emph{IEEE Transactions on Applied Superconductivity}.}

\begin{abstract}

Superconducting parametric amplifiers have great promise for quantum-limited readout of superconducting qubits and detectors. Until recently, most superconducting parametric amplifiers had been based on resonant structures, limiting their bandwidth and dynamic range. Broadband traveling-wave parametric amplifiers based both on the nonlinear kinetic inductance of superconducting thin films and on Josephson junctions are in development. By modifying the dispersion property of the amplifier circuit, referred to as dispersion engineering, the gain can be greatly enhanced and the size can be reduced. We present two theoretical frameworks for analyzing and understanding such parametric amplifiers: (1) generalized coupled-mode equations and (2) a finite difference time domain (FDTD) model combined with a small signal analysis. We show how these analytical and numerical tools may be used to understand device performance.

\end{abstract}

\author{Saptarshi Chaudhuri} \thanks{Corresponding author: schaudh2@stanford.edu}
\affiliation{Department of Physics, Stanford University, Stanford, CA 94305}
\author{Jiansong Gao}
\affiliation{National Institute of Standards and Technology, Boulder, CO 80305}
\author{Kent Irwin}
\affiliation{Department of Physics, Stanford University, Stanford, CA 94305}

\maketitle

\section{Introduction}
Over the past decade, much attention has been focused on the subject of superconducting parametric amplifiers. \cite{beltran},\cite{obrien},\cite{mutus} These amplifiers can have quantum-limited noise performance, as has been demonstrated with Josephson parametric amplifiers that utilize the nonlinear inductance of Josephson junctions. Josephson junction parametric amplifiers have been broadly utilized in the readout and control of superconducting qubits and optomechanical elements. \cite{shankar}, \cite{regal} However, because of the resonant architecture, the gain-bandwidth product of such amplifiers is limited. Much larger bandwidths are necessary for applications such as the multiplexed readout of photon detectors and qubits. \cite{eom}, \cite{macklin} Traveling-wave parametric amplifiers (hereafter referred to as TWPAs) naturally permit large gain-bandwidth products. \cite{yurke} However, in order to achieve high gain over a broad bandwidth, it is necessary to suppress higher harmonic generation and phase match the pump, signal, and idler. \cite{landauer}, \cite{agrawal} In 2010, the dispersion-engineering concept \cite{mohebbi} was first introduced in TWPAs based on the nonlinear kinetic inductance of superconducting thin films; since then, high gain over a broadband has been successfully demonstrated in such amplifiers. \cite{eom}, \cite{bockstiegel} Recently, dispersion-engineered TWPAs based on Josephson junctions have also been proposed. \cite{obrien}, \cite{chaudhurigao}

Here we present new theoretical frameworks for understanding TWPAs. Though we focus on the dispersion-engineered kinetic inductance TWPAs \cite{eom}, the methods we introduce can readily be applied to other TWPA architectures and other nonlinear devices. First, we present new perspectives on the coupled-mode equations, a model that has been adapted from nonlinear optics for understanding TWPAs in the microwave regime. \cite{agrawal} In particular, we have generalized this model to include sidebands of pump harmonics and to demonstrate a computationally efficient scheme for calculating their effect. We then present a second, more flexible model: a finite difference time-domain (FDTD) analysis combined with a small signal computational method, which has certain advantages over the coupled-mode model. Finally, we demonstrate preliminary results for comparing these models to the experimental data.


\section{Kinetic Inductance TWPA and Dispersion Engineering}

In the geometry presented in \cite{eom}, the kinetic inductance TWPA is in the form of a long coplanar waveguide (CPW) patterned from niobium titanium nitride (NbTiN) thin film. In such films, the distributed inductance is dominated by the kinetic inductance, which shows quadratic nonlinearity with current: 
\begin{equation}\label{eq:NKI}
L(I)=L_{0} \left(1 + \frac{I^{2}}{I_{*}^{2}} \right)
\end{equation}
Here $I_{*}$ sets the strength of the nonlinearity. To suppress higher harmonic generation and achieve phase matching, the dispersion property of the transmission line is ``engineered" via the periodic loading process displayed in Fig. 1a. The center strip width of the CPW is periodically increased (impedance decreased), with a physical separation of one-half wavelength corresponding to a preselected frequency $f_{per}$. Much like an electronic bandgap, this creates a stopband (Fig. 1b)--a range of frequencies that are not transmitted by the transmission line--centered at each multiple of $f_{per}$. \cite{eom} Additionally, every third loading is modified to one-half the length of the first two loadings. The modification of every third loading creates weaker stopbands at $f_{per}/3$ and $2f_{per}/3$ and results in modification of the dispersion near these frequencies, as shown in Fig. 1c. By placing the pump frequency near $f_{per}/3$ or $2f_{per}/3$, we can place the third harmonic in a stopband. This limits higher harmonic generation and shock front formation. More importantly, the added dispersion at the pump compensates for the nonlinear phase slippage to meet the phase-matching condition, which enables high gain over a broad bandwidth. 

\begin{figure}
\centerline{\includegraphics[width=16cm]{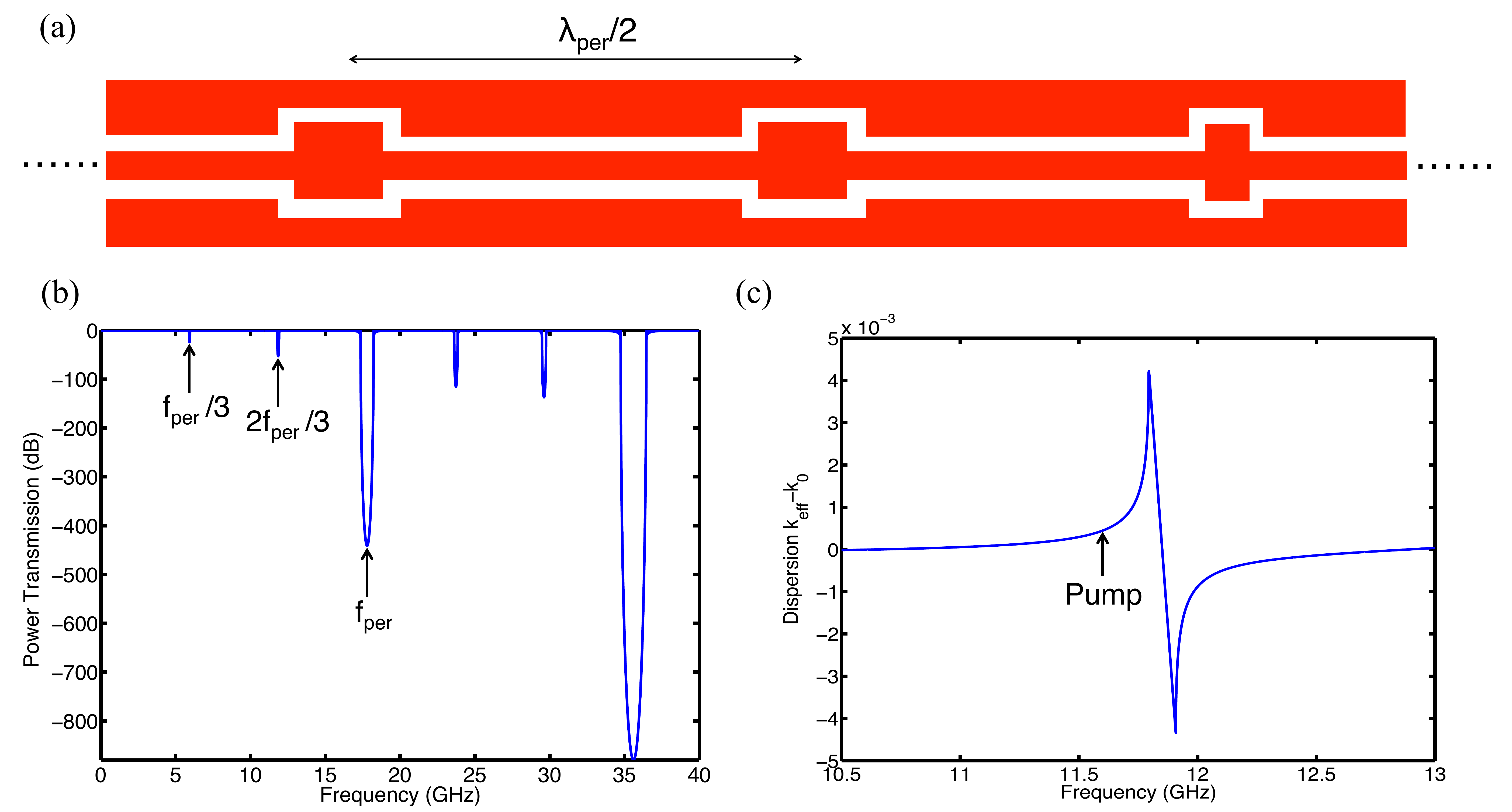}}
\caption{(a) A schematic of the CPW, showing the periodic loading process. (b) The stopband structure created by the loading scheme for the device discussed in Section IV. This device is produced by cascading 224 unit cells (of the form shown in panel (a)), resulting in a transmission line of length 0.6 meters. The impedance of the narrow sections is approximately 300 ohms, while that of the wide sections is 150 ohms. The amplifier is connected to the 50 ohm external circuitry by on-chip adiabatic impedance transformers on the input and output. (c) The dispersion near the second stopband centered at 11.8 GHz, showing the pump placed close to the stopband for phase-matching.}
\end{figure}


\section{Models of Dispersion Engineering}


\subsection{Coupled-Mode Equations}

In this model, we treat the periodic loading as modifying the intrinsic dispersion of the transmission line. The dispersion relation $k(\omega)$ can be computed by assuming translational symmetry and Bloch wave propagation in the periodically loaded transmission line. The wave equation for current in the transmission line is
\begin{equation}\label{eq:NWE}
\frac{\partial^{2}I}{\partial z^{2}} -\frac{\partial}{\partial t} \left[ L(I)C \frac{\partial I}{\partial t} \right]=0
\end{equation}
The solution to the current $I$ can be expressed as the sum of a number of frequency components:
\begin{equation}\label{eq:ICMEdec}
I=\frac{1}{2} \left(\sum_{\alpha} A_{\alpha}(z)\mathrm{exp}(i(\omega_{\alpha}t-k_{\alpha}z)) + c.c. \right)
\end{equation}
where c.c. represents the complex conjugate of the first term and the summation is over index $\alpha$. 

The $I^{2}dI/dt$ term in equation (\ref{eq:NWE}) connects combinations of four frequencies. Thus, if a pump at frequency $\omega_{p}$ and signal at frequency $\omega_{s}$ are input into the transmission line, sum and difference frequencies of the form $n\omega_{p}+m\omega_{s}$, where $m$ and $n$ are integers, will be generated; furthermore, given the cubic form of $I^{2}dI/dt$, $m+n$ is constrained to be odd.  Since we wish to develop a coupled-mode analysis appropriate for the periodically loaded kinetic inductance TWPA, and the dispersion engineering strongly suppresses the generation of pump harmonics, we can safely set the amplitude of pump harmonics to zero. Also, we assume that the signal is much weaker than the pump, so all frequencies with $m>1$, corresponding to second- and higher-order sidebands, can be ignored. We can index the first-order sideband frequencies by $\omega_{n}=n\omega_{p}+\omega_{s}-\omega_{p}$, where $n$ runs from $-\infty$ to $\infty$, their respective wavenumbers by $k_{n}$, and their respective amplitudes by $A_{n}(z)$. In this notation, the sidebands of the pump harmonic $n\omega_{p}$ are represented by the frequencies $\omega_{n}$ and $\omega_{-n}$. (See Fig. 2a.) The cubic form of the $I^{2}dI/dt$ nonlinearity implies that we need only consider $n$ odd.

Furthermore, we assume that the signal gain and sideband generation is small enough that the pump current is not significantly depleted. That is, we assume that the pump current amplitude is a constant. Using these approximations and the slow-wave approximation, we derive the pump amplitude $A_{p}(z)$, and  a set of differential equations for the sideband amplitudes $A_{n}(z)$:
\begin{equation}\label{eq:CMEp}
A_{p}(z)=|A_{p}(0)|\mathrm{exp}(-i\Delta\phi z)
\end{equation}
\begin{equation}\label{eq:CMEs}
\frac{dA_{n}}{dz}=-i\frac{\omega_{n}}{\omega_{p}} \Delta\phi \left[A_{n-2}\mathrm{exp}(i(\Delta k_{n} - 2\Delta\phi)z) + 2A_{n} + A_{n+2}\mathrm{exp}(-i(\Delta k_{n+2}- 2\Delta\phi)z) \right]
\end{equation}
where
\begin{equation}\label{eq:SPM}
\Delta\phi=\frac{k_{p}}{8I_{*}^{2}}|A_{p}(0)|^{2}
\end{equation}
is the pump self-phase modulation per unit length that arises from the AC pump current interacting with itself and
\begin{equation}\label{eq:deltak}
\Delta k_{n}=k_{n}-k_{n-2}-2k_{p}
\end{equation}
is the linear phase mismatch among the pump tone, the sideband at $\omega_{n}$, and the sideband at $\omega_{n-2}$. Equation (\ref{eq:CMEs}), in contrast to previous works \cite{obrien}, \cite{eom}, and \cite{agrawal}, incorporates all sidebands. We have assumed $A_{p}(0)=|A_{p}(0)|$, for computational convenience. In practice, we limit the computation to the sidebands of the first $J$ harmonics. We utilize the transformation
\begin{equation}\label{eq:AntoBn}
B_{n}(z)=A_{n}(z)\mathrm{exp} \left(i \left(\frac{2\omega_{n}}{\omega_{p}} \Delta\phi -\frac{1}{2}(\vec{v}_{n} \cdot \vec{K}) \right) z \right)
\end{equation}
where the vector $\vec{v}_{n}$ is a J+1-component vector defined component-wise by
\begin{equation}\label{eq:vn}
(\vec{v}_{n})_{j}=
\begin{cases}
+1, & 1\leq j \leq 1+(n+J)/2 \\
-1, &  1+(n+J)/2 < j \leq J+1
\end{cases}
\end{equation}
and $\vec{K}=(0, \Delta k_{-J+2} + 2\Delta\phi,  \Delta k_{-J+4} + 2\Delta\phi, ... ,  \Delta k_{J} + 2\Delta\phi)$. Under this transformation, combining (\ref{eq:CMEs}), (\ref{eq:AntoBn}), and (\ref{eq:vn}), we arrive at
\begin{equation}\label{eq:CME-lin}
\frac{dB_{n}}{dz}=-i\frac{\omega_{n}}{\omega_{p}} \Delta\phi B_{n-2} - \frac{i}{2} \left(\vec{v}_{n} \cdot \vec{K} \right) B_{n} -i\frac{\omega_{n}}{\omega_{p}} \Delta\phi B_{n+2}
\end{equation}
The advantage of (\ref{eq:CME-lin}) over (\ref{eq:CMEs}) is that the coefficients of the amplitudes are now independent of position. This allows us to use matrix eigenvalue-eigenvector methods instead of numerical differential equation solvers to determine the sideband amplitudes, dramatically reducing the computation time. 

Coupled-mode analyses to date have considered only the signal and idler frequencies, corresponding to $J=1$. \cite{obrien}, \cite{agrawal} In this case, from $(\ref{eq:CME-lin})$ we can derive a closed form expression for the signal power gain of a transmission line of length $L$: 
\begin{equation}\label{eq:Gsgen}
G_{s}=\left| \frac{A_{1}(L)}{A_{1}(0)} \right|^{2} = \left| \mathrm{cosh}(gL) - \frac{i(\Delta k_{1}+2\Delta\phi)}{2g} \mathrm{sinh}(gL) \right|^{2}
\end{equation}
where
\begin{equation}\label{eq:pgc}
g=\sqrt{\frac{\omega_{s}\omega_{i}}{\omega_{p}^{2}}(\Delta\phi)^{2}- \left(\frac{\Delta k_{1}+2\Delta\phi}{2}\right)^{2}}
\end{equation}
is the parametric gain coefficient. For a dispersionless line, in which $\Delta k_{1}=0$, the above expression for gain reduces to, for signal frequencies near the pump,
\begin{equation}\label{eq:Gsquad}
G_{s}\approx 1+(L\Delta\phi)^{2}
\end{equation}
The gain varies quadratically with line length and pump power. The maximum gain is achieved when the phases of the pump, signal, and idler are matched. Phase-matching occurs when $\Delta k_{1}+2\Delta\phi=0$, in which case (\ref{eq:Gsgen}) reduces to
\begin{equation}\label{eq:Gsexp}
G_{s} \approx 1+\mathrm{sinh}^{2}(L\Delta\phi)
\end{equation}
for $\omega_{s} \approx \omega_{p}$. The gain varies exponentially with line length and pump power. More generally, (\ref{eq:Gsgen}) and (\ref{eq:pgc}) show that, for a fixed line length $L$, the closer $\Delta k_{1}+2\Delta\phi$ is to zero, the higher the gain will be. In this sense, $\Delta k_{1}+2\Delta\phi$ is a measure of the phase mismatch. As illustrated in Fig. 1, the periodic loading process increases the wavenumber at the pump, while having negligible effect on the wavenumber of the signal and idler. As such, the phase mismatch $\Delta k_{1}+2\Delta\phi$ is reduced as compared to that in dispersionless lines and the gain-bandwidth product is greatly enhanced. Note that equations (\ref{eq:Gsgen})-(\ref{eq:Gsexp}) only hold as long as the pump is much stronger than the signal.

The gain calculation corresponding to the cases $J=1$ and $J=3$ are compared to experimental data later in this paper. We include the $J=3$ model because third harmonic sidebands directly mix with the signal and idler and thus may have a significant effect on the signal amplitude. The inclusion of these higher harmonic sidebands is therefore expected to yield a more reliable analysis. 

\begin{figure}
\centerline{\includegraphics[width=16cm]{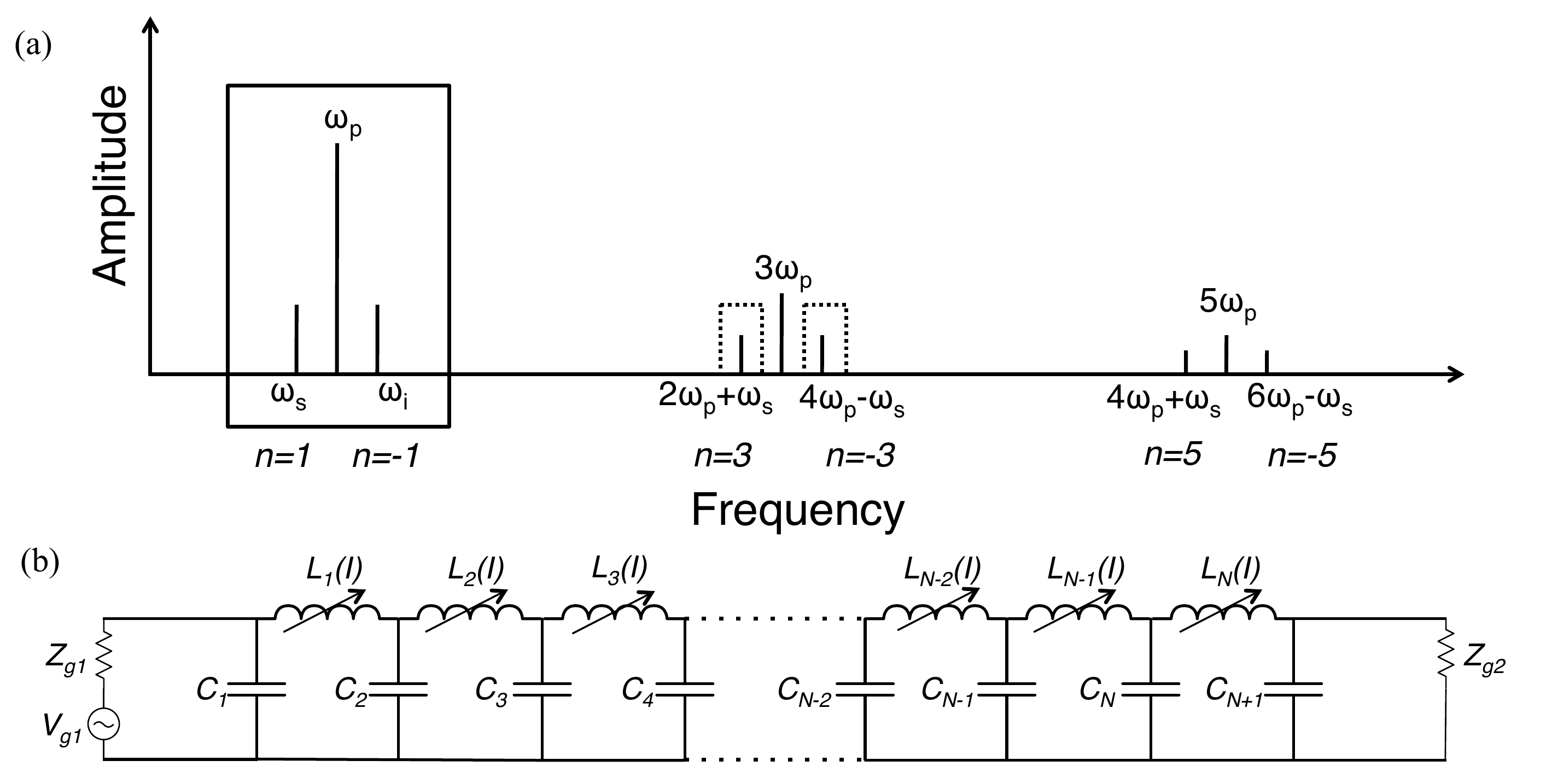}}
\caption{(a) Explanation of sidebands and coupled-mode analysis. The solid box contains the tones included in a traditional coupled-mode analysis ($J=1$). The dashed boxes represent the first two harmonic sidebands that can be included in our generalized analysis. (b) The circuit used in the FDTD method.}
\end{figure}


\subsection{Finite-Difference Time Domain and Small Signal Method}

The coupled-mode equations provide a simple method for understanding dispersion-engineered TWPAs, but the model has a few drawbacks. The assumptions of translational symmetry and intrinsic dispersion are not valid near the two ends of a transmission line of finite length. The periodic changes in impedance produce reflections and standing waves. This results in variations in the pump current, and if the variation is not sufficiently small, the assumption of constant pump current is not accurate. In the above coupled-mode analysis, we have made the simplification to not include pump harmonics. In addition, we would like an independent way to verify the coupled-mode model.
For these reasons, we have developed a finite-difference time domain method that is more flexible and general. The discretized circuit is shown in Fig. 2b. The line has a total of $N$ inductors and $N+1$ capacitors. We can write differential equations for the current $I_{r}$ through the $r$th inductor and the voltage $V_{r}$ across $r$th capacitor.
\begin{equation}\label{eq:FDTDdI}
V_{r+1}-V_{r}=-L_{r} \left(1+ \frac{I_{r}^{2}}{I_{*r}^{2}} \right) \frac{dI_{r}}{dt}
\end{equation}
for $1\leq r \leq N$ and
\begin{equation}\label{eq:FDTDdV}
I_{r}-I_{r-1}=-C_{r}\frac{dV_{r}}{dt}
\end{equation}
for $1\leq r \leq N+1$. Here, $I_{0}$ and $I_{N+1}$ are the input and output current in the circuit and are related to the boundary conditions by
\begin{equation}\label{eq:BDC1}
V_{g1}-I_{0}Z_{g1}=V_{1}
\end{equation}
\begin{equation}\label{eq:BDC2}
V_{N+1}-I_{N+1}Z_{g2}=0
\end{equation}
We allow each inductor to have its own $L_{r}$, $C_{r}$ and $I_{*r}$, to account for an arbitrary loading pattern.

One possible FDTD method to calculate the signal gain would be to numerically solve the above differential equations including both a pump and a signal in the drive. However, in this approach, the simulation must be conducted at every signal frequency, which is computationally intensive for broadband gain.

We have developed a more efficient way to calculate the broadband signal gain. In this approach, we perform the simulation in two steps. First, we solve for the pump and harmonic amplitudes using FDTD and then Fast Fourier Transform (FFT); this is a nonlinear problem. Then, since the signal is weak, we perturbatively solve for the signal and sidebands using the pump amplitudes derived in the first step, which reduces to a linear problem that can be solved rapidly with matrix methods. In this manner, we only utilize numerical differential equation solvers once for the entire calculation, instead of once for each signal frequency, which results in a faster algorithm.

Thus, we first solve for the pump waveform traveling on the LC ladder in the absence of the signal. Setting the voltage drive $V_{g1}(t)=V_{p}\mathrm{sin}(\omega_{p}t)$, we can solve for the steady-state of (\ref{eq:FDTDdI})-(\ref{eq:BDC2}) using numerical differential equation solvers. We then perform a Fast Fourier Transform (FFT) to determine the amplitude of the pump current harmonics in each unit cell.

Next, we determine the evolution of an input signal along the line. In a similar manner to that utilized in coupled-mode, we can write the currents as a sum of pump harmonic components and sidebands components:
\begin{equation}\label{eq:IFDTDdec}
I_{r}(t)=\frac{1}{2} \left(\sum_{n=1}^{\infty} I_{r}(n\omega_{p})\mathrm{exp}(in\omega_{p}t) + \sum_{n=1}^{\infty} I_{r}(\omega_{n})\mathrm{exp}(i\omega_{n}t) + c.c. \right)
\end{equation}
and similarly for the voltage across each capacitor. Plugging (\ref{eq:IFDTDdec}) into (\ref{eq:FDTDdI})-(\ref{eq:BDC2}) yields linearized equations for the sideband amplitudes. Assuming a signal drive voltage of $V_{s}\mathrm{sin}(\omega_{s}t)$, we find, for $1\leq r \leq N$, 
\begin{equation}\label{eq:SSA1}
\frac{I_{r}(\omega_{n})-I_{r+1}(\omega_{n})}{C_{r+1}} - \frac{I_{r-1}(\omega_{n})-I_{r}(\omega_{n})}{C_{r}}=\omega_{n}^{2}L_{r} \left[I_{r}(\omega_{n}) + \frac{1}{4I_{*r}^{2}} \sum_{k,l,m}I_{r}(k\omega_{p})I_{r}(l\omega_{p})I_{r}(\omega_{m})\delta_{n,k+l+m} \right]
\end{equation}
\begin{equation}\label{eq:SSA2}
\frac{I_{0}(\omega_{n})-I_{1}(\omega_{n})}{C_{1}}-\omega_{s}V_{s}\delta_{n,1}+i\omega_{n}I_{0}(\omega_{n})Z_{g1}=0
\end{equation}
\begin{equation}\label{eq:SSA3}
\frac{I_{N+1}(\omega_{n})-I_{N}(\omega_{n})}{C_{N+1}}+i\omega_{n}I_{N+1}(\omega_{n})Z_{g2}=0
\end{equation}
In practice, we truncate the calculation to a finite number of pump harmonics and sidebands. (\ref{eq:SSA1})-(\ref{eq:SSA3}) can then be rearranged into a matrix equation, and the amplitudes $I_{r}(\omega_{n})$ solved; from this, we may extract the signal gain. In this manner, we need only solve for the pump and harmonics once, resulting in a fast and efficient algorithm.


\section{Results and Discussion}

In this section, we show how the above models may be used to provide insight on device performance. We compare model predictions to gain data from the test device displayed in Fig. 1a. We first compute the expected impedance and phase velocity of the narrow and wide CPW sections. We adjust these parameters to approximately match the observed width and center frequency of the first and second stopbands. We utilize three models: (1) coupled-mode analysis with only signal and idler, (2) coupled-mode analysis with both first and third harmonic sidebands, and (3) time-domain simulation with the first three pump harmonics and the sidebands of these harmonics used in the small signal calculation. 

The observed gain and the gain curves predicted by our models are shown in Fig. 3. The pump frequency for the test device is 11.561 GHz. The large ripple in the observed gain is likely due to an impedance mismatch between the amplifier and the external circuitry. \cite{eom} The smoothed gain curve is plotted for guiding purposes. For the coupled-mode analysis with both first and third harmonic sidebands ($J=3$), we adjust the pump current until the gain matches the observed gain near the pump. We use the same pump current for the coupled-mode analysis with only the signal and idler ($J=1$). For the time-domain analysis, we split the transmission line into approximately 130 LC unit cells per pump wavelength. The value of $I_{*}$ is taken to be a factor of three larger in the wide CPW sections compared to the narrow sections. We adjust the pump voltage to match the gain near the pump. 

For the time-domain method, the computation time was dominated by solving for the pump and harmonics using differential equation solvers. In contrast, the computation time for coupled-mode was significantly shorter. Thus, as will be discussed, it is desirable to be able to solve for the pump and harmonics in a perturbative frequency-domain model, just as we presently do for the signal and sidebands.

As illustrated, all three models reproduce the broad bandwidth observed in experiment, with gain above 10 dB over a bandwidth of approximately 3.5 GHz. Note the two dips in the gain near the pump, corresponding to the signal and idler falling in the stopband.

We also note that the coupled-mode model with only signal and idler predicts slightly higher gain than that including the third harmonic sidebands, especially at frequencies detuned far from the pump. At these frequencies, mixing between the signal and third harmonic sidebands can be comparable to the mixing between the signal and idler, as the latter process is more poorly phase-matched (relative to frequencies near the pump). At least at these frequencies, the gain calculation should include higher order sidebands (e.g. use the $J=3$ model). The small difference between the gain of the $J=3$ model and the time-domain model shows that the presence of third pump harmonic does not significantly affect gain, which is expected because the harmonic lies in a stopband. 

\begin{figure}
\centerline{\includegraphics[width=16cm]{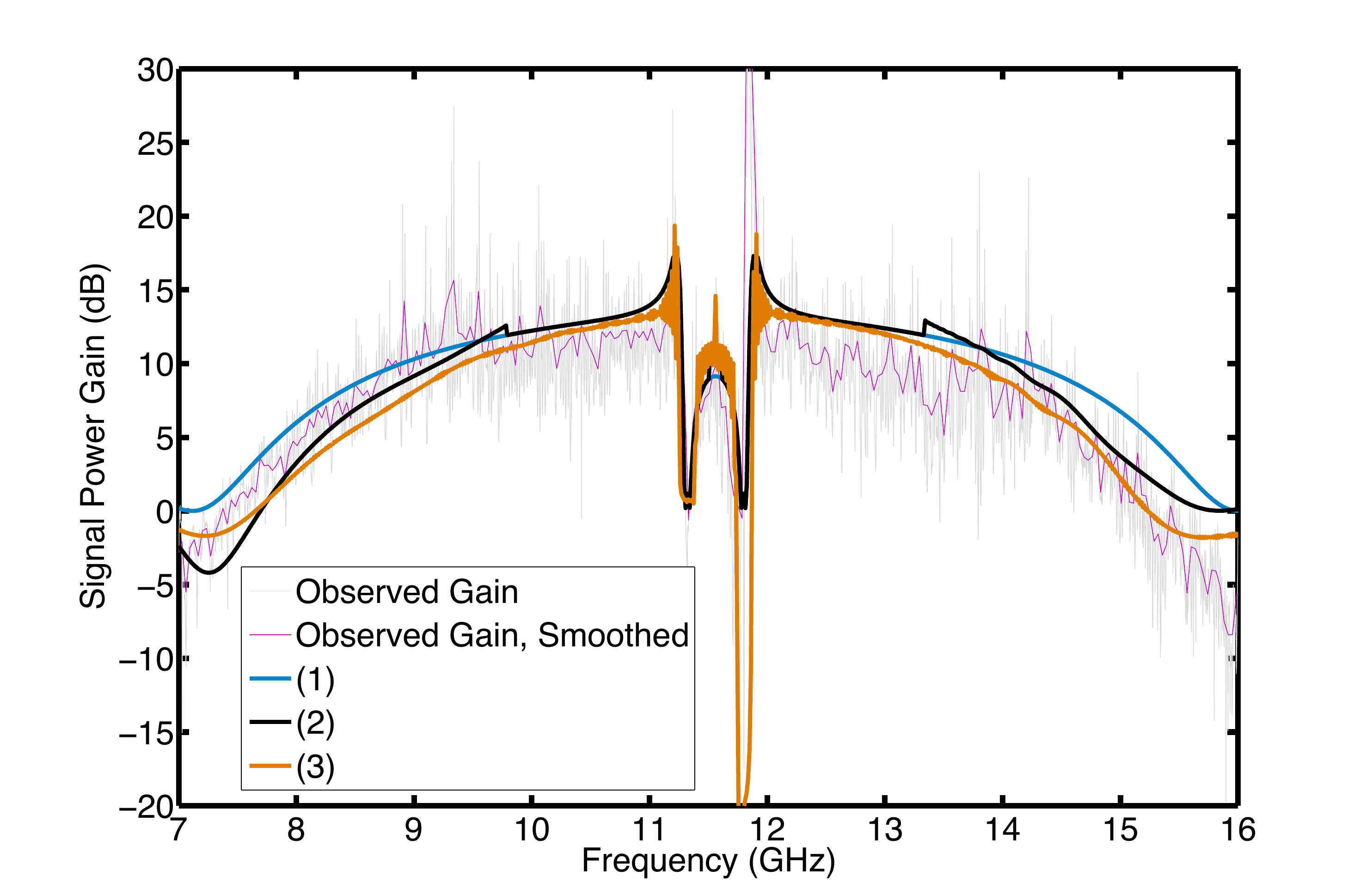}}
\caption{Observed and predicted gain profiles. The numbers in the legend correspond to the three different models listed in the text: (1) coupled-mode analysis with only signal and idler, (2) coupled-mode analysis with both first and third harmonic sidebands, and (3) time-domain simulation with the first three pump harmonics and the sidebands of these harmonics used in the linearized signal calculation. The discontinuities in the gain profile of (2) near 9.75 GHz and 13.35 GHz are the result of setting to zero the amplitude of any third harmonic sidebands falling in the stopband.}
\end{figure}


\section{Conclusion}

We have demonstrated two theoretical frameworks for understanding dispersion-engineered TWPAs. First, we have shown how the traditional coupled-mode analyses may be efficiently extended to take into account sidebands of higher harmonics. We have also developed an FDTD method, combined with a small signal analysis, which has certain advantages over the coupled-mode model. We have demonstrated how these models may be used to understand device performance. We have focused on the application of these algorithms to dispersion-engineered TWPAs based on kinetic inductance, but the algorithms may readily be applied to other TWPAs, such as those based on Josephson junctions. Further work will include using the models to study the effects of sidebands of pump harmonics on signal gain, using the models to optimize the dispersion engineering design to achieve higher gain and bandwidth, and extending these models to discuss dissipation mechanisms and amplifier noise. We have also developed a harmonic balance computation scheme in which the pump, in addition to the signal, is solved in the frequency domain using perturbation theory. This scheme combines the fast computation times of a coupled-mode analysis with the flexibility of the time-domain method. \cite{chaudhuri} Together, these tools represent a powerful method for understanding dispersion engineering in TWPAs.

\section{Acknowledgment}
The authors would like to thank B.H. Eom and P.K. Day for providing the experimental data used in Section IV. We would also like to thank J. Zmuidzinas for useful discussions. We acknowledge support from the NASA APRA grant \#NNX13AQ98G. Author S. Chaudhuri acknowledges NASA NSTRF support.

\end{document}